\documentclass[pre,twocolumn,amsmath,amssymb]{revtex4}
\usepackage{amsmath,amssymb,latexsym}
\usepackage{graphicx,epstopdf}
\allowdisplaybreaks[1]

\newcommand{\bs}[1]{\boldsymbol{#1}}
\newcommand{\ca}[1]{\mathcal{#1}}
\newcommand{\wO}{$\ca{O}$ }
\newcommand{\wOe}{$\ca{O}$}
\newcommand{\wL}{$\ca{L}_{\theta}$ }
\newcommand{\wLe}{$\ca{L}_{\theta}$}
\newcommand{\widthfig}{0.45\textwidth}
\newcommand{\vspfigA}{\vspace{1eM}}  
\newcommand{\vspfigB}{\vspace{0eM}} 
\newcommand{\vspfigC}{\vspace{1eM}} 

\begin{document}
	
\title{Thermodynamic transports in a circular system with a temperature difference} 


\author{
Tooru Taniguchi, Colin Bain McRae, and  Shin-ichi Sawada
}
 
\affiliation{
School of Science and Technology, Kwansei Gakuin University, 2-1 Gakuen, Sanda, Hyogo, Japan
}

\date{\today}

\begin{abstract} 

Thermodynamic transport phenomena in the system consisting of many hard-disks confined in a circular tube with a temperature difference are discussed. 
Here, temperatures on parts of the walls of the tube are imposed by stochastic boundary conditions for particles to contact with these thermal walls. 
In this system, the temperature difference induces, not only energy currents, but also a circulating particle current, inside the tube. 
Transport properties of these steady currents are discussed in various values of system parameters, such as the temperature difference, the particle density, the width of the tube, and the positions of the thermal walls. 

\end{abstract}
   
\pacs{
05.60.Cd, 
05.70.Ln, 
45.50.Jf 
}

\maketitle  
\section{Introduction}

Thermodynamic transports, in which thermodynamic quantities such as temperature, chemical potential, and particle density play an essential role, appear in a wide variety of natural phenomena. 
A typical example of such transport phenomena is heat currents, as energy currents caused by temperature gradients. 
This type of energy transport is described macroscopically by nonequilibrium thermodynamics  \cite{GM84,D14}, especially by Fourier's law near equilibrium states. 
For the heat currents, many works have already been done to justify and/or generalize their phenomenological descriptions in microscopic viewpoints \cite{LL03,D08}.  
Another type of thermodynamic transport is particle currents caused by gradients of chemical potential or particle density \cite{GM84,D14}. 
This type of current has many important applications, such as electric currents \cite{D95,I97}.  
The B\'{e}nard convention is another example of the thermodynamic particle transports, and is caused by a gravitational force and a temperature difference between the upper side and the lower side of the system \cite{MW06,L10}. 
One may also mention the ratchet systems generating particle currents caused by asymmetric potentials and thermally random forces \cite{AN07,K10}.  

The principal aim of this paper is to discuss thermodynamic transports by many particles confined in a circular tube attached to two heat reservoirs.  
As concrete particles we use hard disks, which have been widely used to discuss many-particle effects because of a simplicity of their dynamics \cite{AT87,H92,M08,HM13}. 
The two heat reservoirs induce different temperatures in specific parts of walls (thermal walls) of the circular tube. 
The effect of temperatures on the thermal walls is introduced as the thermal boundary conditions which are represented as collision rules of hard disks with the thermal walls \cite{CT80,MK84,BL91,TT98,TS17}.    
We show that in this system a circulating particle current is caused by the temperature difference and disk-disk collisions. 
This particle current occurs without any external force acting on particles in the direction of the current inside the tube,  differently from the B\'{e}nard convention and the ratchet systems. 
We discuss properties of the direction and the magnitude of this steady particle current inside the circular tube. 

The temperature difference of the thermal walls also induces energy currents inside the tube. 
It is noted that this circular system has two different routes to transfer energy from a low-temperature area to a high-temperature area, and is regarded as a thermal network system. 
From this viewpoint, we discuss how the energy currents via these different routes depend on various system parameters, such as the temperature difference, the particle density, the positions of the thermal walls, and the width of the tube. 
Especially, the energy currents, as well as the circulating particle current, are investigated in a wide range of the temperature difference, showing properties beyond its linear responses. 
Magnitudes of these currents are varied, not only by bulk properties of the system (e.g. disk-disk interactions) and the temperature difference, but also by the spatial geometry on the circular tube and the positions of the thermal walls.  

This paper is organized as follows. 
In Sect. \ref{HardDisksCircularTubeAttachedTwoHeatReservoirs}, we introduce the system consisting of many hard-disks in a circular tube with thermal walls with different temperatures. 
In Sects. \ref{CirculatingParticleCurrents} and \ref{EnergyCurrents}, we discuss a circulating particle current and energy currents  inside the tube, respectively, and system parameter dependences of those currents. 
Finally, we give conclusions and remarks in Sect. \ref{ConclusionRemarks}.

\section{Hard disks in a circular tube with a temperature difference}
\label{HardDisksCircularTubeAttachedTwoHeatReservoirs} 

\begin{figure}[!t]
\vspfigA
\begin{center}
\includegraphics[width=\widthfig]{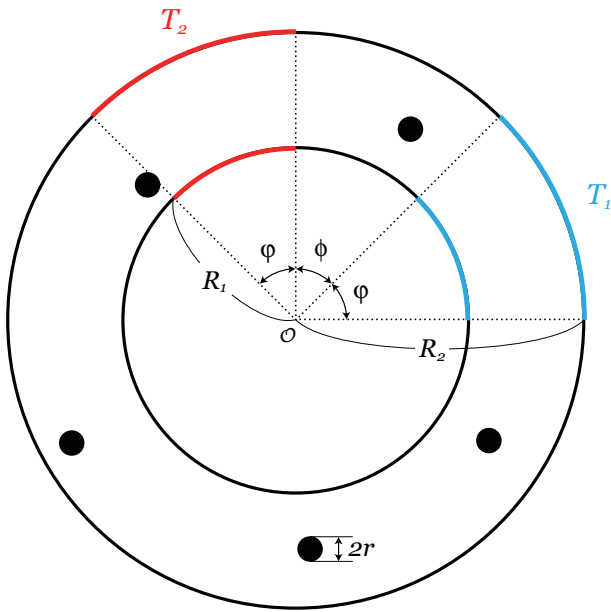}
\vspfigB
\vspace{1eM}
\caption{
A schematic illustration of the system consisting of hard disks with the radius $r$ in a circular tube attached to two heat reservoirs. 
The circular tube is the area enclosed by the circle wall with the radius $R_{1}$ and the one with the radius $R_{2}$ ($> R_{1} + 2r$) with the same center \wOe. 
With the angle $\theta$ of counterclockwise rotation around the origin \wOe, the blue (red) parts in $0\leq\theta<\varphi$ ($\varphi + \phi\leq\theta< 2\varphi + \phi$) as the thermal wall $W_{1}$ ($W_{2}$) are attached to a heat reservoir of the temperature $T_{1}$ ($T_{2}$). 
} 
\label{Fig1System}
\end{center}
\vspfigC
\end{figure}

The system, which we consider in this paper, consists of hard disks confined in a circular tube attached to two heat reservoirs, as shown in Fig. \ref{Fig1System}. 
Here, the number of disks is $N$, each hard disk has the mass $m$ and the radius $r$, and the circular tube is given as the area enclosed by the circle wall with the radius $R_{1}$ and the one with the radius $R_{2}$ ($>R_{1} + 2r$) with the same center \wOe. 
With the angle $\theta$ of counterclockwise rotation around the origin \wOe, the part of the circle walls in $0\leq\theta<\varphi$ ($\varphi + \phi\leq\theta<2\varphi+\phi$) with positive angles $\varphi \in (0,\pi]$ and $\phi\in[0,2(\pi-\varphi)]$ as the thermal wall $W_{1}$ ($W_{2}$) shown by the blue (red) lines in Fig. \ref{Fig1System}, are attached to a heat reservoir with the temperature $T_{1}$ ($T_{2}$). 
%

We describe effects of the thermal walls $W_{1}$ and $W_{2}$, whose temperatures are $T_{1}$ and $T_{2}$, respectively, as stochastic boundary conditions. 
Under these conditions, the probability density function $f_{j}(\bs{p})$ of the momentum $\bs{p}$ of a hard disk just after colliding with the thermal wall $W_{j}$ at the position $\bs{q}$ is given by  
\begin{eqnarray}
   f_{j}(\bs{p}) = \frac{\bs{n}(\bs{q})\cdot\bs{p}}{\left(2\pi\right)^{1/2}\left(mk_{B}T_{j}\right)^{3/2}} 
   \exp\left(-\frac{\left|\bs{p}\right|^{2}}{2mk_{B}T_{j}}\right) 
\label{InjecMomenDistr1}
\end{eqnarray}
with the Boltzmann constant $k_{B}$, independently of the momentum of the disk just before colliding the thermal wall \cite{CT80,MK84,BL91,TT98,TS17}. 
Here, $\bs{n}(\bs{q})$ is the unit vector which is directed toward the inside of the circular tube from the colliding position $\bs{q}$ of the disk and is perpendicular to the wall $W_{j}$ at the position $\bs{q}$. 
[Note that $f_{j}(\bs{p})$ is the probability density function of $\bs{p}$ for $0<\bs{n}(\bs{q})\cdot\bs{p}<+\infty$ and $-\infty< \bs{n}^{\prime}(\bs{q})\cdot\bs{p}<+\infty$ for the unit vector $\bs{n}^{\prime}(\bs{q})$ orthogonal to $\bs{n}(\bs{q})$.] 
Each hard disk collides elastically with the walls other than these thermal walls $W_{1}$ and $W_{2}$, as well as with other hard disks.  
We assume that the momentum change in each collision of disks with walls or other disks occur instantly, and there is no instant change of the disk position in each collision. 
Except in such disk-wall or disk-disk collisions, each hard disk moves with a constant velocity.

In the following sections, we discuss transport properties in steady states of the system by numerical calculations.  
In these calculations, we use the unit of $m=1$, $k_{B}T_{1} = 1$, and $2R_{2} = 1$, and also take $N=100$.
We use the particle density $\rho \equiv N r^{2}/(R_{2}^{2}-R_{1}^{2})$ to specify the disk radius $r$ for given values of $\rho$, $N$, $R_{1}$, and $R_{2}$. 
In order to discuss the transport properties, we calculate various time-average quantities and distributions based on data over the time-interval $\ca{T} = 10^{7}$, except for the data for Figs. \ref{Fig12ParCurEneRatio}, \ref{Fig13ParCurDenHigh}, \ref{Fig14KinetEnerg}, and the inset of Fig. \ref{Fig6EneCurAB} with $\ca{T} = 2\times 10^{6}$. 
We take the time-interval $\ca{T}$ after calculating particle orbits at an early stage so that in this time-interval the system can be regarded to be in a steady state. 


\section{Circulating particle currents}
\label{CirculatingParticleCurrents}

In this section, we discuss the average particle current $I$ via a cross section \wL of the circular tube. 
Here, the cross section \wL is the line of intersection of the circular tube area and a semi-infinite straight line drawn from the origin \wO with the angle $\theta$. 
With the number $N_{+}(\tau)$ ($N_{-}(\tau)$) of disks passing through the cross section \wL in the positive (negative) angle direction in the time interval $\tau$, the particle current $I$ is defined as $\lim_{\tau\rightarrow +\infty} [N_{+}(\tau)-N_{-}(\tau)]/\tau$ in a steady state, although we calculate it numerically as the quantity $[N_{+}(\ca{T})-N_{-}(\ca{T})]/\ca{T}$ approximately. 
The particle current $I$ is independent of the angle $\theta$ of \wL in the steady state because of the conservation of number of disks.

\begin{figure}[!t]
\vspfigA
\begin{center}
\includegraphics[width=\widthfig]{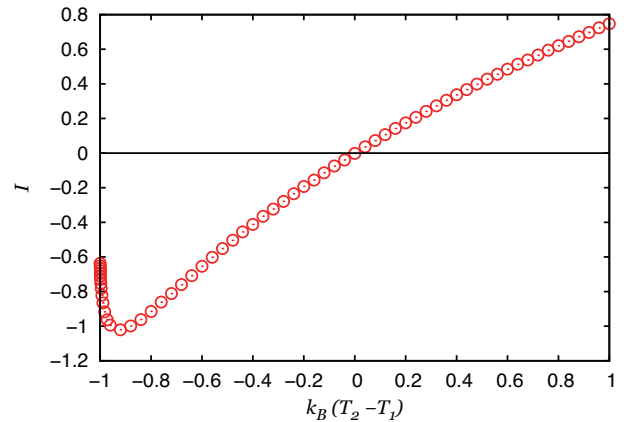}
\vspfigB
\caption{
The particle current $I$ as a function of $k_{B}(T_{2}-T_{1})$ in the case of $\rho = 0.01$, $R_{1}=0.3$, and $\varphi = \phi = \pi/4$.
Here, and in all figures hereafter, we use the dimensionless units of $m=1$, $k_{B}T_{1} = 1$, and $2R_{2} = 1$. 
}
\label{Fig2ParCurTem}
\end{center}
\vspfigC
\end{figure}  
%
In Fig. \ref{Fig2ParCurTem} we show the graph of the particle current $I$ as a function of $k_{B}(T_{2}-T_{1}) \in (-1,1]$ in the case of $\rho = 0.01$ (so $r=0.004$), $R_{1}=0.3$, and $\varphi = \phi = \pi/4$. 
(Taking the value $k_{B}T_{1} =1$, in this graph we change the value of $k_{B}T_{2}$ in the range of $(0,2]$.)
This figure shows that a non-zero particle current $I$ is caused inside the tube by a difference of the temperatures of the two heat reservoirs attached to the system.  
If the two temperatures $T_{1}$ and $T_{2}$ coincide, then the current $I$ disappears, because the system is in an equilibrium state. 
The direction of the current $I$ is clockwise for $T_{2}<T_{1}$ and counterclockwise for $T_{2}>T_{1}$ in the tube shown in Fig. \ref{Fig1System}, and is a increasing function of $k_{B}(T_{2}-T_{1})$ except in very small values of $k_{B}T_{2}$. 
In the cases of very small $k_{B}T_{2}$, the magnitude of the particle current $I$ is suppressed, as shown in cases of $k_{B}(T_{2}-T_{1})$ to be very close to $-1$ in Fig. \ref{Fig2ParCurTem}, probably because speeds of disks can become extremely slow just after colliding the disks with the thermal wall $W_{2}$ with a very low temperature $T_{2}$.

\begin{figure}[!t]
\vspfigA
\begin{center}
\includegraphics[width=\widthfig]{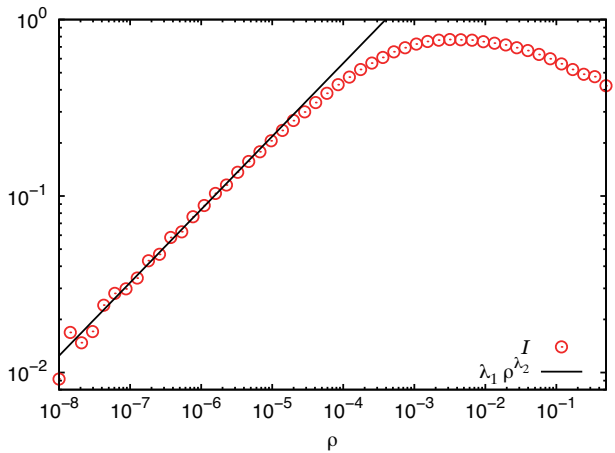}
\vspfigB
\caption{
The particle current $I$ (red circles) as a function of the particle density $\rho$ as a log-log plot in the case of $k_{B}T_{2}=2$, $R_{1}=0.3$, and $\varphi = \phi = \pi/4$. 
The black line is a fit of the current $I$ in a low-density region to the function $\lambda_{1}\rho^{\lambda_{2}}$ of $\rho$ with the fitting parameters $\lambda_{1}$ and $\lambda_{2}$. 
}
\label{Fig3ParCurDen}
\end{center}
\vspfigC
\end{figure}  
%
Figure \ref{Fig3ParCurDen} shows the particle current $I$ (red circles) as a function of the particle density $\rho$ as a log-log plot in the case of $k_{B}T_{2}=2$, $R_{1}=0.3$, and $\varphi = \phi = \pi/4$. 
It is indicated in this figure that the magnitude of the current $I$ is suppressed in a high-density region as well as in a low-density region, and takes a peak in a middle density.   
It may be noted that the momentum distribution $f_{j}(\bs{p})$ for disks just after colliding with the thermal wall $W_{j}$, defined in Eq. (\ref{InjecMomenDistr1}), is invariant under the change of sign of the momentum component in the direction of the angle $\theta$ ($j=1,2$), so particles just after colliding with a thermal wall do not have a non-zero average particle current in the $\theta$-direction. 
In addition, collisions of disks with the hard-walls other than the thermal walls do not change the $\theta$-components of momenta of disks. 
Therefore, we suppose that in the low-density limit $\rho\rightarrow 0$, i.e. in the ideal-gas case without any particle-particle interaction, the particle current $I$ would disappear. 
Figure \ref{Fig3ParCurDen} supports this argument, showing that the particle current $I$ monotonically decreases as the particle density $\rho$ decreases in a low-density region. 
To discuss this point more quantitatively, we added in Fig. \ref{Fig3ParCurDen} a fit (black line) of the current $I$ in a low-density region to the power function $\lambda_{1}\rho^{\lambda_{2}}$ of $\rho$ with the fitting parameters $\lambda_{1}$ and $\lambda_{2}$. 
The particle current $I$ in the low-density region is nicely fitted to this power function with the values $\lambda_{1} = 25.7$ and $\lambda_{2} = 0.414$, showing a slow decay of the current $I$ for the particle density $\rho$ to decrease. 
%

\begin{figure}[!t]
\vspfigA
\begin{center}
\includegraphics[width=\widthfig]{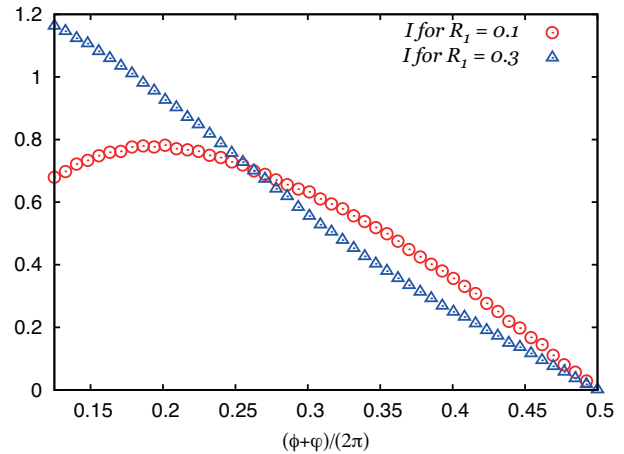}
\vspfigB
\caption{
The particle currents $I$ as functions of $(\varphi+\phi)/(2\pi)$ in the cases of $R_{1}=0.1$ (red circles), $R_{1}=0.3$ (blue triangles), $k_{B}T_{2}=2$, $\rho = 0.01$, and $\varphi = \pi/4$.
}
\label{Fig4ParCurAng}
\end{center}
\vspfigC
\end{figure}  
%
In order to discuss how the particle current $I$ depends on the positions of the two kinds of thermal walls, we show the graph of the currents $I$ as functions of $(\varphi+\phi)/(2\pi)$ in the cases of $R_{1}=0.1$ (red circles), $R_{1}=0.3$ (blue triangles), $k_{B}T_{2}=2$, $\rho = 0.01$, and $\varphi = \pi/4$, in Fig. \ref{Fig4ParCurAng}. 
Here, we fix the angle $\varphi$ as $\varphi = \pi/4$, so that the starting angle $\varphi+\phi$ of the thermal wall $W_{2}$ must be within the range $[\pi/4,7\pi/4)$. 
By the mirror inversion of the system in a line passing the origin \wO with the angle $\varphi/2$, the current $I|_{\varphi + \phi = \theta}$ is transformed into $-I|_{\varphi + \phi = 2\pi-\theta}$, so that the particle current $I$ satisfies the relation $I|_{\varphi + \phi = \theta} = -I|_{\varphi + \phi = 2\pi-\theta}$ for any angle $\theta\in[\pi/4,7\pi/4)$.  
Noting this point, we plotted the graph of $I$ for $\varphi+\phi \in [\pi/4,\pi/2]$ in Fig. \ref{Fig4ParCurAng}. 
This relation leads to $I|_{\varphi + \phi = \pi} =0$, as also shown in this figure, by the mirror symmetry of the system in the line passing the origin \wO with the angle $\varphi/2$. 
Figure \ref{Fig4ParCurAng} shows that the particle current $I$ is a monotonically decreasing function of $\varphi + \phi$ in the case of $R_{1}=0.3$, although the current $I$ take a maximum value in a middle angle of $\varphi + \phi$ in the case of $R_{1}=0.1$. 
In any case, an asymmetry between the two routes connecting the two thermal walls $W_{1}$ and $W_{2}$, i.e. between the area in $\varphi<\theta<\varphi+\phi$ and the area in $\varphi+\phi<\theta<2\pi$, is essential to cause the circulating particle current $I$.

\begin{figure}[!t]
\vspfigA
\begin{center}
\includegraphics[width=\widthfig]{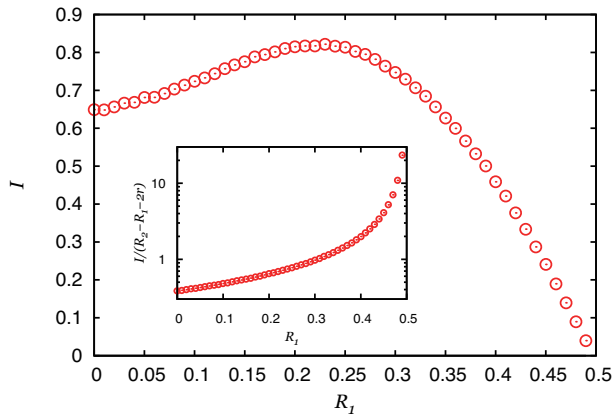}
\vspfigB
\caption{
The particle current $I$ in the main figure, and the particle current density $I/(R_{2}-R_{1}-2r)$ as a linear-log plot in the inset, as a function of the inner radius $R_{1}$ of the circular tube in the case of $k_{B}T_{2}=2$, $\rho = 0.01$, and $\varphi= \phi = \pi/4$. 
}
\label{Fig5ParCurCen}
\end{center}
\vspfigC
\end{figure}  
%
As another property of the particle current $I$, we discuss how the current $I$ depends on the inner radius $R_{1}$ of the tube under a fixed particle-density $\rho$. 
For this property, we show Fig. \ref{Fig5ParCurCen} for the graph of the particle current $I$ in the main figure, and the graph of the particle current density $I/(R_{2}-R_{1}-2r)$ as a linear-log plot in the inset, as a function of the inner radius $R_{1}$ of the circular tube in the case of $k_{B}T_{2}=2$, $\rho = 0.01$, and $\varphi= \phi = \pi/4$. 
(Noting that the centers of hard disks pass the cross section \wL of the tube via the line of the length $R_{2}-R_{1}-2r$, we define the particle current density by the current $I$ divided by this length.)  
This figure indicates that the magnitude of the current $I$ takes its maximum value around the middle of the range $(0,R_{2})$ of the variable $R_{1}$, while the particle current density $I/(R_{2}-R_{1}-2r)$ is a monotonically increasing function of $R_{1} \in (0,R_{2})$ differently from the current $I$ itself.

\section{Energy currents}
\label{EnergyCurrents}

The model introduced in Sec. \ref{HardDisksCircularTubeAttachedTwoHeatReservoirs} can also be regarded as a network model of energy currents caused by a temperature difference of the attached two heat reservoirs. 
One of the important characteristics in this network model is that there are two ways of carrying energy from the hot thermal wall to the cold thermal wall via the positive and negative angle directions inside the tube. 
Now, we proceed to discuss these energy currents in this circular-tube model.

In order to discuss such energy currents concretely, we introduce the energy current $J(\theta)$ passing the cross section \wL of the circular tube with the angle $\theta$ in a steady state. 
This energy current occurs by the two kinds of particle motions; one is by the passages of particles through the cross section \wLe, and the other is by the collisions of two disks located on the different sides of the cross section \wLe. 
Noting this fact we introduce the quantity $K_{+}(\tau,\theta)$ ($K_{-}(\tau,\theta)$) as the sum of the kinetic energy of particles passing through the cross section \wL in the positive (negative) direction of the angle $\theta$ in the time interval $\tau$. 
We also introduce the quantity $K_{c}(\tau,\theta)$ as the sum of changes of kinetic energies of particles in the side of the positive angle direction from the cross section \wLe, by colliding with the particles in the side of the negative angle direction from the cross section \wLe, in the time interval $\tau$. 
Using these quantities $K_{+}(\tau,\theta)$, $K_{-}(\tau,\theta)$, and $K_{c}(\tau,\theta)$, the energy current $J(\theta)$ is given by $J(\theta)=\lim_{\tau\rightarrow+\infty}[K_{+}(\tau,\theta)-K_{-}(\tau,\theta) + K_{c}(\tau,\theta)]/\tau$. 
(In actual numerical simulations, we calculate the current $J(\theta)$ by $[K_{+}(\ca{T},\theta)-K_{-}(\ca{T},\theta) + K_{c}(\ca{T},\theta)]/\ca{T}$ approximately.)

Because of the energy conservation in particle collisions except in those with the thermal walls, the energy current $J(\theta)$ is constant for the angle $\theta \in [\varphi,\varphi+\phi)$ or $\theta \in [2\varphi+\phi,2\pi)$. 
Based on this fact, we introduce the quantities $J_{A}$ and $J_{B}$ as the energy currents $-J(\theta)$ in $\theta\in [\varphi,\varphi+\phi)$ and  $J(\theta)$ in $\theta\in [2\varphi+\phi,2\pi)$, respectively.
In other words, the current $J_{A}$ ($J_{B}$) is the energy current flowing in the negative (positive) angle direction from the thermal wall $W_{2}$ to the thermal wall $W_{1}$.

As another type of energy current, we can also consider the energy current $\ca{J}_{j}$ injected into the system from the heat reservoir with the temperature $T_{j}$ in a steady state ($j=1,2$).  
This current is given by $\ca{J}_{j} = \lim_{\tau\rightarrow+\infty}\ca{K}_{j}(\tau)/\tau $, where $\ca{K}_{j}(\tau)$ is the sum of changes of kinetic energies of particles by their collisions with the thermal wall $W_{j}$ in the time interval $\tau$. 
(In actual numerical simulations, we calculate the current $\ca{J}_{j}$ by $\ca{K}_{j}(\ca{T})/\ca{T}$ approximately.)  
The energy currents $\ca{J}_{1}$, $\ca{J}_{2}$, $J_{A}$, and $J_{B}$ satisfy the relations 
\begin{eqnarray}
   \ca{J}_{1} &=& - \ca{J}_{2},  \label{RelatJ12} \\
   \ca{J}_{2} &=& J_{A} + J_{B} \label{RelatJ2AB}
\end{eqnarray}
by the equations of continuity for the particle energy. 
Equation (\ref{RelatJ12}) simply means that the energy injected into the system must be equal to the energy removed from the system in a steady state. 
On the other hand, Eq. (\ref{RelatJ2AB}) means that the energy current $\ca{J}_{2}$ via the thermal wall $W_{2}$ is divided into the two currents $J_{A}$ and $J_{B}$ flowing via different routes connecting two thermal walls.   
In our numerical calculations, the energy currents $\ca{J}_{1}$, $\ca{J}_{2}$, $J_{A}$, and $J_{B}$ were calculated independently, and were checked to satisfy Eqs. (\ref{RelatJ12}) and (\ref{RelatJ2AB}) in some concrete cases, although we omit to show them in this paper.

\begin{figure}[!t]
\vspfigA
\begin{center}
\includegraphics[width=\widthfig]{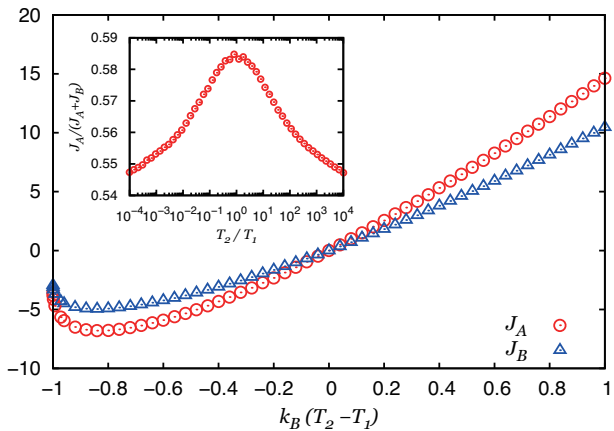}
\vspfigB
\caption{
The energy currents $J_{A}$ (red circles) and $J_{B}$ (blue triangles) as functions of $k_{B}(T_{2}-T_{1})$  in the main figure, and the ratio $J_{A}/(J_{A}+J_{B})$ of the current $J_{A}$ to the total energy current $J_{A}+J_{B}$ as a function of $T_{2}/T_{1}$ for $T_{2}/T_{1}\neq 1$ as a log-linear plot in the inset, in the case of $\rho = 0.01$, $R_{1}=0.3$, and $\varphi = \phi = \pi/4$. 
}
\label{Fig6EneCurAB}
\end{center}
\vspfigC
\end{figure}  
%
As the main figure of Fig. \ref{Fig6EneCurAB} we show the graphs of the energy currents $J_{A}$ (red circles) and $J_{B}$ (blue triangles) as functions of $k_{B}(T_{2}-T_{1})$ in the case of $\rho = 0.01$, $R_{1}=0.3$, and $\varphi = \phi = \pi/4$. 
This figure shows that the energy currents $J_{A}$ and $J_{B}$ flow from the hot thermal wall to the cold thermal wall. 
Besides, in this case, the magnitude $|J_{A}|$ of the current $J_{A}$ flowing the shorter route of the tube connecting the thermal walls $W_{1}$ and $W_{2}$ is larger than the magnitude $|J_{B}|$ of the current $J_{B}$ flowing the longer route of the tube in the case of $T_{1} \neq T_{2}$. 
It is also indicated in this figure that the magnitudes $|J_{A}|$ and $|J_{B}|$ of these currents increase for the absolute value $|T_{2}-T_{1}|$ of the temperature difference to increase, except in very small values of the temperature $T_{2}$ near the case of $k_{B}(T_{2}-T_{1}) = -1$. 
In the inset of Fig. \ref{Fig6EneCurAB} we also show the graph of the ratio $J_{A}/(J_{A}+J_{B})$ of the energy current $J_{A}$ to the total energy current $J_{A}+J_{B} (=J_{2})$ as a function of $T_{2}/T_{1}$ for $T_{2}/T_{1} \neq 1$ as a log-linear plot in the case of $\rho = 0.01$, $R_{1}=0.3$, and $\varphi = \phi = \pi/4$. 
(For this graph, we change the value of $k_{B}T_{2}$ in the wider range of $[10^{-4},10^{4}]$ than in the main figure of Fig. \ref{Fig6EneCurAB} with $k_{B}T_{2} \in [10^{-4},2]$.) 
This graph shows that the ratio of the current $J_{A}$ to the total energy current $J_{A}+J_{B}$ decreases as the absolute value of the logarithm of the temperature ratio $T_{2}/T_{1}$ increases (so as the absolute value $|T_{1}-T_{2}|$ of the temperature difference increases) in fixed positions of the thermal walls $W_{1}$ and $W_{2}$. 
It also suggests that the ratio $J_{A}/(J_{A}+J_{B})$ is symmetric under the change of the sign of $\ln (T_{2}/T_{1})$.

\begin{figure}[!t]
\vspfigA
\begin{center}
\includegraphics[width=\widthfig]{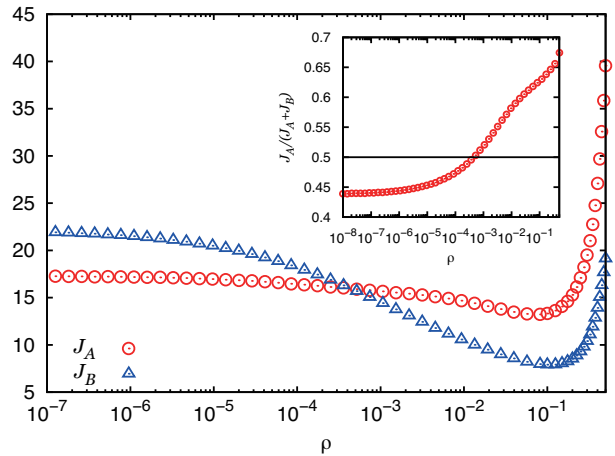}
\vspfigB
\caption{
The energy currents $J_{A}$ (red circles) and $J_{B}$ (blue triangles) in the main figure, and the ratio $J_{A}/(J_{A}+J_{B})$ of the current $J_{A}$ to the total energy current $J_{A}+J_{B}$ in the inset, as functions of the particle density $\rho$ as log-linear plots in the case of $k_{B}T_{2}=2$, $R_{1}=0.3$, and $\varphi = \phi = \pi/4$.
}
\label{Fig7EneCurDen}
\end{center}
\vspfigC
\end{figure}  
%
It is important to note that the magnitude of the energy current flowing the shorter route of the tube is not always larger than the one flowing the longer route of the tube. 
In order to discuss this point, we show Fig. \ref{Fig7EneCurDen} for the graphs of the energy currents $J_{A}$ (red circles) and $J_{B}$ (blue triangles) in its main figure, and the graph of the ratio $J_{A}/(J_{A}+J_{B})$ of the current $J_{A}$ to the total energy current $J_{A}+J_{B}$ in its inset, as functions of the particle density $\rho$, as log-linear plots, in the case of $k_{B}T_{2}=2$, $R_{1}=0.3$, and $\varphi = \phi = \pi/4$. 
This figure shows that the magnitude of the energy current $J_{A}$ flowing the shorter route of the tube is smaller than the one of the current $J_{B}$ in cases of low particle-densities, while in cases of high particle-densities the current $J_{A}$ is larger than the current $J_{B}$. 
Besides, the main figure of Fig. \ref{Fig7EneCurDen} suggests that the energy currents $J_{A}$ and $J_{B}$ have their minimum values around $\rho = 0.1$, and increase rapidly as $\rho$ increases in high-density cases. 
Furthermore, the inset of Fig. \ref{Fig7EneCurDen} indicates that the ratio $J_{A}/(J_{A}+J_{B})$ (crossing the line of $1/2$ showing the relation of $J_{A} = J_{B}$ at a middle value of $\rho$) is a monotonically increasing function of $\rho$.

\begin{figure}[!t]
\vspfigA
\begin{center}
\includegraphics[width=\widthfig]{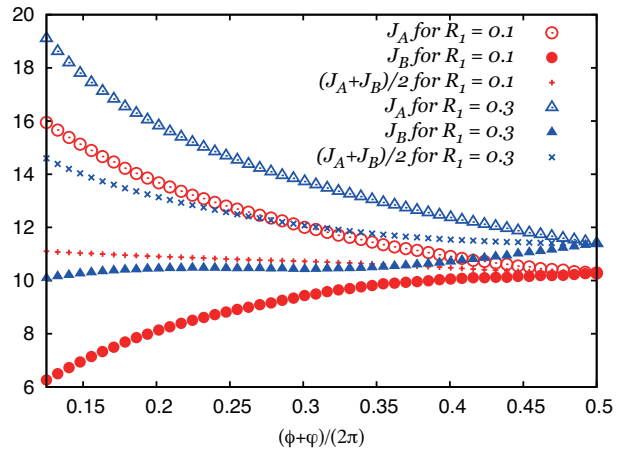}
\vspfigB
\caption{
The energy currents $J_{A}$ (red circles), $J_{B}$ (red filled circles), and their average $(J_{A}+J_{B})/2$ (red pluses) for $R_{1}=0.1$, and the energy currents $J_{A}$ (blue triangles), $J_{B}$ (blue filled triangles), and their average $(J_{A}+J_{B})/2$ (blue crosses) for $R_{1}=0.3$, as functions of $(\varphi+\phi)/(2\pi)$ in the case of $k_{B}T_{2}=2$, $\rho = 0.01$, and $\varphi = \pi/4$. 
}
\label{Fig8EneCurAng1-2}
\end{center}
\vspfigC
\end{figure}  
%
In Fig. \ref{Fig8EneCurAng1-2} we show the graphs of the energy currents $J_{A}$ (red circles), $J_{B}$ (red filled circles), and their average $(J_{A}+J_{B})/2$ (red pluses) for $R_{1}=0.1$, and the graphs of the energy currents $J_{A}$ (blue triangles), $J_{B}$ (blue filled triangles), and their average $(J_{A}+J_{B})/2$ (blue crosses) for $R_{1}=0.3$, as functions of $(\varphi+\phi)/(2\pi)$ in the case of $k_{B}T_{2}=2$, $\rho = 0.01$, and $\varphi = \pi/4$. 
By the mirror inversion of the system in a line passing the origin \wO with the angle $\varphi/2$, the current $J_{A}|_{\varphi+\phi = \theta}$ is transformed into $J_{B}|_{\varphi+\phi = 2\pi - \theta}$, so that the energy currents satisfies the relation $J_{A}|_{\varphi+\phi = \theta} = J_{B}|_{\varphi+\phi = 2\pi - \theta}$. 
Noting this point we plotted the energy currents for $\varphi+\phi \leq \pi$ in Fig. \ref{Fig8EneCurAng1-2}. 
It is shown in this figure that in the case of $\rho = 0.01$ the magnitude of the energy currents $J_{A}$ and $J_{B}$ become larger in the shorter route of the tube. 
This figure also indicates that the total energy current $J_{A} + J_{B}$ from the hot thermal wall to the cold thermal wall, as twice the quantity $(J_{A}+J_{B})/2$ shown in Fig. \ref{Fig8EneCurAng1-2}, is a decreasing function of $\varphi+\phi$ for $\varphi+\phi \leq \pi$, taking its minimum value at $\varphi+\phi = \pi$ in which the hot and cold thermal walls are located on the exact opposite sides of the circular tube with each other and the particle current $I$ disappears. 
The energy currents $J_{A}$ and $J_{B}$ in the case of $R_{1}=0.3$ are larger than the currents $J_{A}$ and $J_{B}$ in the case of $R_{1}=0.1$, respectively.  


\begin{figure}[!t]
\vspfigA
\begin{center}
\includegraphics[width=\widthfig]{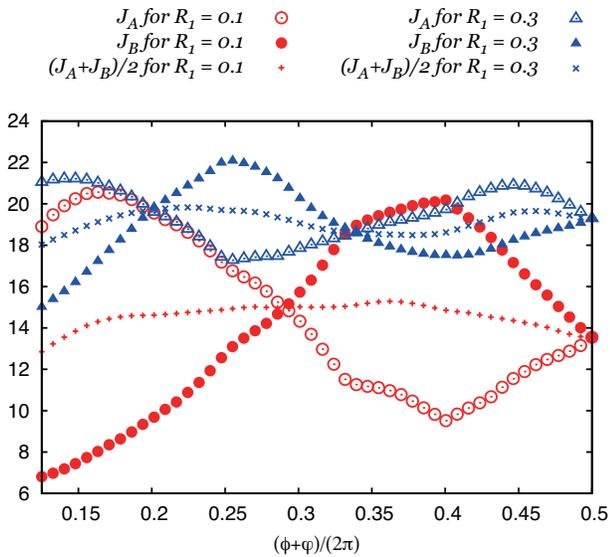}
\vspfigB
\caption{
The energy currents $J_{A}$ (red circles), $J_{B}$ (red filled circles), and their average $(J_{A}+J_{B})/2$ (red pluses) for $R_{1}=0.1$, and the energy currents $J_{A}$ (blue triangles), $J_{B}$ (blue filled triangles), and their average $(J_{A}+J_{B})/2$ (blue crosses) for $R_{1}=0.3$, as functions of $(\varphi+\phi)/(2\pi)$ in the case of $k_{B}T_{2}=2$, $\rho = 10^{-8}$, and $\varphi = \pi/4$.
}
\label{Fig9EneCurAng1-8}
\end{center}
\vspfigC
\end{figure}  
%
In order to discuss effects of particle-particle interactions in the energy currents $J_{A}$ and $J_{B}$ we consider rather much lower density cases than in Fig. \ref{Fig8EneCurAng1-2}. 
In Fig. \ref{Fig9EneCurAng1-8} we show the graphs of the energy currents $J_{A}$ (red circles), $J_{B}$ (red filled circles), and their average $(J_{A}+J_{B})/2$ (red pluses) for $R_{1}=0.1$, and the graphs of the energy currents $J_{A}$ (blue triangles), $J_{B}$ (blue filled triangles), and their average $(J_{A}+J_{B})/2$ (blue crosses) for $R_{1} = 0.3$, as functions of $(\varphi+\phi)/(2\pi)$ in the case of $k_{B}T_{2}=2$, and $\varphi = \pi/4$  in the extremely low density $\rho = 10^{-8}$.  
This figure shows that in this extremely low density the energy currents $J_{A}$ and $J_{B}$ oscillate as functions of the angle $\varphi+\phi$, and the oscillating period is smaller in a narrower tube. 
Besides, in each of these cases, there are the multiple angles in which the currents $J_{A}$ and $J_{B}$ coincide with each other, and the total energy current $J_{A} + J_{B}$ is not a simple decreasing function of $\varphi+\phi$ for $\varphi+\phi \leq \pi$. 
Since effects of particle-particle interactions should be extremely small in the case of $\rho = 10^{-8}$, we suppose that these oscillatory behaviors of the energy currents $J_{A}$ and $J_{B}$ are caused by a geometrical shape of the system including the positions of the thermal walls. 
It also implies that monotonically increasing or decreasing properties of the energy currents  $J_{A}$,  $J_{B}$, and $(J_{A}+J_{B})/2$ as functions of $(\varphi+\phi)/(2\pi)$ in Fig. \ref{Fig8EneCurAng1-2} are caused by disk-disk collisions.

\begin{figure}[!t]
\vspfigA
\begin{center}
\includegraphics[width=\widthfig]{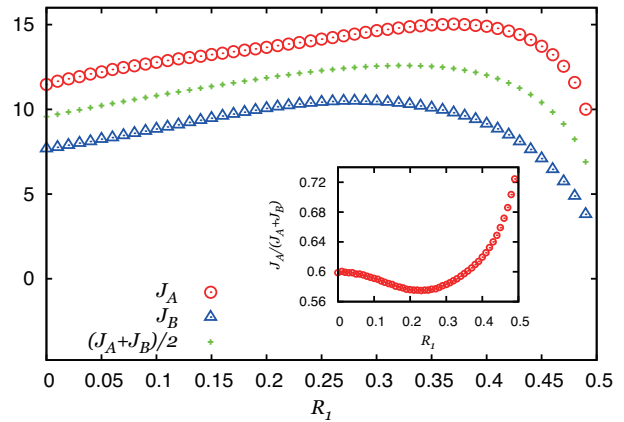}
\vspfigB
\caption{
The energy currents $J_{A}$ (red circles), $J_{B}$ (blue triangles), and their average $(J_{A}+J_{B})/2$ (green pluses) in the main figure, and the ratio $J_{A}/(J_{A}+J_{B})$ of the current $J_{A}$ to the total current $J_{A}+J_{B}$ in the inset, as functions of the inner radius $R_{1}$ of the circular tube in the case of $k_{B}T_{2}=2$, $\rho = 0.01$, and $\varphi= \phi = \pi/4$. } 
\label{Fig10EneCurCen1-2}
\end{center}
\vspfigC
\end{figure}  
%
In Fig. \ref{Fig10EneCurCen1-2} we show the graphs of the energy currents $J_{A}$ (red circles) and $J_{B}$ (blue triangles), and their average $(J_{A}+J_{B})/2$ (green pluses) in the main figure, and the graph of the ratio $J_{A}/(J_{A}+J_{B})$ of the current $J_{A}$ to the total current $J_{A}+J_{B}$  in the inset, as functions of the inner radius $R_{1}$ of the circular tube in the case of $k_{B}T_{2}=2$, $\rho = 0.01$, and $\varphi= \phi = \pi/4$. 
This figure shows that not only the energy currents $J_{A}$ and $J_{B}$ but also the current ratio $J_{A}/(J_{A}+J_{B})$ depend on the inner radius $R_{1}$ of the circular tube in the case of a fixed particle-density and fixed angles $\varphi$ and $\phi$ for the thermal walls. 
The main figure of Fig. \ref{Fig10EneCurCen1-2} indicates that each of the energy currents $J_{A}$ and $J_{B}$, as well as their average $(J_{A}+J_{B})/2$, as a function of $R_{1}$ takes its maximum value around the middle of the range $(0,R_{2})$ of $R_{1}$, although the values of $R_{1}$ at their maximum values are different with each other. 
On the other hand, it is shown in the inset of Fig. \ref{Fig10EneCurCen1-2} that the current ratio $J_{A}/(J_{A}+J_{B})$  takes its minimum value around the middle of the range $(0,R_{2})$ of $R_{1}$.

\begin{figure}[!t]
\vspfigA
\begin{center}
\includegraphics[width=\widthfig]{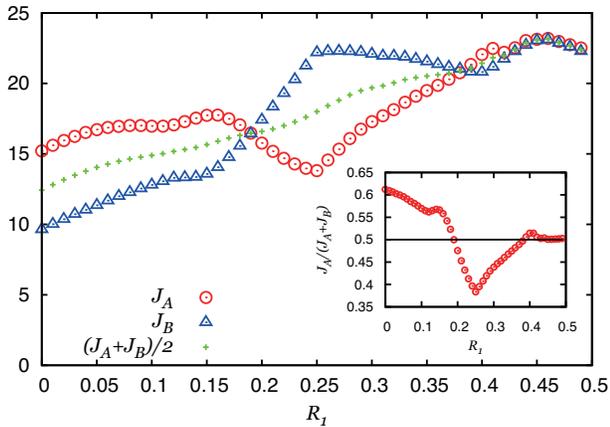}
\vspfigB
\caption{
The energy currents $J_{A}$ (red circles), $J_{B}$ (blue triangles), and their average $(J_{A}+J_{B})/2$ (green pluses) in the main figure, and the ratio $J_{A}/(J_{A}+J_{B})$ of the current $J_{A}$ to the total current $J_{A}+J_{B}$ in the inset, as functions of the inner radius $R_{1}$ of the circular tube in the case of $k_{B}T_{2}=2$, $\rho = 10^{-8}$, and $\varphi= \phi = \pi/4$. 
} 
\label{Fig11EneCurCen1-8}
\end{center}
\vspfigC
\end{figure}  
%
We show Fig. \ref{Fig11EneCurCen1-8} for the graphs of the energy currents $J_{A}$ (red circles), $J_{B}$ (blue triangles), and their average $(J_{A}+J_{B})/2$ (green pluses) in the main figure, and the graph of the ratio $J_{A}/(J_{A}+J_{B})$ of the current $J_{A}$ to the total current $J_{A}+J_{B}$ in the inset, as functions of the inner radius $R_{1}$ of the circular tube in the case of $k_{B}T_{2}=2$, $\rho = 10^{-8}$, and $\varphi= \phi = \pi/4$. 
In contrast to the case of $\rho = 0.01$ shown in Fig. \ref{Fig10EneCurCen1-2}, the main figure of Fig. \ref{Fig11EneCurCen1-8} for an extremely low-density case of $\rho = 10^{-8}$ shows oscillatory behaviors in the energy currents $J_{A}$ and $J_{B}$ as functions of $R_{1}$. 
(In other words, the absence of such oscillatory behaviors in the energy currents in Fig. \ref{Fig10EneCurCen1-2} would be regarded as an effect of disk-disk collisions.)
The inset of Fig. \ref{Fig11EneCurCen1-8} shows that the ratio $J_{A}/(J_{A}+J_{B})$ of the current $J_{A}$ to the total current $J_{A}+J_{B}$ crosses the line of $0.5$ plural times as a function of $R_{1}$, and is close to the value $0.5$ in cases of very narrow tubes. 
The main figure of Fig. \ref{Fig11EneCurCen1-8} also indicates that the total energy current $J_{A}+J_{B}$ increases as the inner radius $R_{1}$ increases except in cases of very narrow tubes.

\section{Conclusions and Remarks}
\label{ConclusionRemarks}

In this paper, we considered thermodynamic transport phenomena caused by many hard-disks in a circular tube attached to two heat reservoirs with different temperatures. 
The attachments of the circular tube to two heat reservoirs were described by thermal collision rules of hard disks with those attached parts of the walls (i.e. thermal walls) of the tube by which the momentum probability distribution of hard disks after colliding with a thermal wall is given from an equilibrium distribution with the temperature of the wall. 
We showed that the temperature difference of the two thermal walls can cause, not only energy currents from the hot thermal wall to the cold thermal wall, but also a circulating particle current inside the tube. 

The circulating particle current $I$, which is the average current of disks passing a cross section of the tube in a steady state, takes a constant value independent of the position of the cross section of the circular tube, and occurs in the direction from the cold thermal wall to the hot thermal wall through the shorter route of the tube connecting these two kinds of thermal walls. 
(Therefore, for this current $I$ it is necessary for the two thermal walls not to be located on the exact opposite sides of the circular tube with each other.)  
Disk-disk collisions would also be essential to cause this particle current $I$, since this current would disappear in the low particle-density limit without any disk-disk collision, i.e. in an ideal gas. 
The magnitude of the particle current $I$ becomes larger for larger temperature differences of the thermal walls except in cases for one of the temperatures of the thermal walls to be extremely small. 
In our result, the particle current $I$ appears as a nonlinear function of the temperature difference.
The current $I$ is suppressed, not only in very low particle-density cases, but also in very high particle-density cases.  
The magnitude of $I$ as a function of the width of the tube under a constant particle-density takes its maximum value at a middle value of the width of the tube.

In the circular-tube system, there exist two different routes connecting the thermal walls $W_{1}$ and $W_{2}$ with different temperatures. 
This feature, as a thermal network system, induces the two kinds of steady energy currents $J_{A}$ and $J_{B}$, each of which flows through each of these two routes, in the direction from the hot thermal wall to the cold thermal wall. 
The magnitudes of the energy currents $J_{A}$ and $J_{B}$ become larger for a larger temperature difference of the thermal walls except in cases for one of the temperatures of the thermal walls to be extremely small.  
We investigated the energy currents $J_{A}$ and $J_{B}$ in a wide range of the temperature difference so that these currents depend on the temperature difference nonlinearly. 
Each of the currents $J_{A}$ and $J_{B}$ as a function of the particle density takes its minimum value at a middle value of the particle density.  
In high particle-density cases with many disk-disk collisions, each of the energy currents $J_{A}$ and $J_{B}$ becomes smaller in the longer route of the tube connecting the thermal walls $W_{1}$ and $W_{2}$. 
In these cases, each of the currents $J_{A}$ and $J_{B}$ as a function of the width of the tube under a constant particle-density takes its maximum value at a middle value of the width of the tube. 
In contrast, in extremely low-density cases, the currents $J_{A}$ and $J_{B}$ show oscillatory behaviors as functions of the thermal wall location and as functions of the width of the tube under a constant particle-density. 
These results indicate especially that the energy currents, as well as the circulating particle current, can be varied not only by bulk properties of the system but also by the spatial geometry of the system. 
Our results also show that the ratio $J_{A}/(J_{A}+J_{B})$ of the current $J_{A}$ to the total current $J_{A}+J_{B}$ depends on the temperature difference, the particle density, the positions of the thermal walls, and the width of the tube.

\begin{figure}[!t]
\vspfigA
\begin{center}
\includegraphics[width=\widthfig]{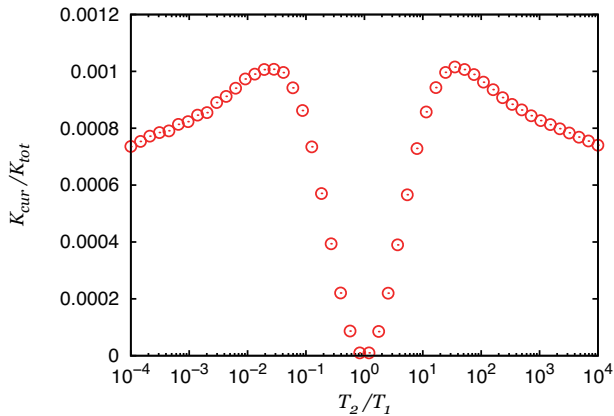}
\vspfigB
\caption{
The kinetic energy ratio $K_{cur}/K_{tot}$ as a function of $T_{2}/T_{1}$, as a log-linear plot, in the case of $\rho = 0.01$, $R_{1}=0.3$, and $\varphi = \phi = \pi/4$. 
}
\label{Fig12ParCurEneRatio}
\end{center}
\vspfigC
\end{figure}  
%
It may be meaningful to estimate what amount of ratio of kinetic energy to the total kinetic energy contributes to the circulating particle current $I$ caused by a temperature difference. 
As momenta related to the particle current $I$, we use the rotating component $p_{\theta}^{(j)}$ of the momentum $\bs{p}^{(j)}$ of the $j$-th disk, with the positive direction of the rotating component $p_{\theta}^{(j)}$ around the origin \wO as same as that of the angle $\theta$, for $j=1,2,\cdots,N$.  
By using the momentum $p_{\theta}^{(j)}$, we introduce the average kinetic energy $K_{cur}\equiv [(1/N)\sum_{j=1}^{N}\langle p_{\theta}^{(j)}\rangle]^{2}/(2m)$ per disk with the time average $\langle p_{\theta}^{(j)}\rangle$ of $p_{\theta}^{(j)}$, contributing directly to the circulating particle current $I$. 
Besides, we define the average total kinetic energy $K_{tot}$ per disk by $K_{tot}\equiv (1/N)\sum_{j=1}^{N}\langle |\bs{p}^{(j)}|^{2}\rangle/(2m)$ with the time average $\langle |\bs{p}^{(j)}|^{2}\rangle$ of $ |\bs{p}^{(j)}|^{2}$. 
With these kinetic energies we introduce the kinetic energy ratio $K_{cur}/K_{tot}$ of the kinetic energy $K_{cur}$ contributing the circulating particle current $I$ to the total kinetic energy $K_{tot}$. 
In Fig. \ref{Fig12ParCurEneRatio} we show the graph of this kinetic energy ratio $K_{cur}/K_{tot}$ as a function of $T_{2}/T_{1}$, as a log-linear plot, in the case of $\rho = 0.01$, $R_{1}=0.3$, and $\varphi = \phi = \pi/4$. 
This figure indicates that the kinetic energy ratio $K_{cur}/K_{tot}$ for the particle current $I$ is only about $10^{-3}$ of the total kinetic energy at most in $T_{2}/T_{1} \in [10^{-4},10^{4}]$. 
This implies that disk movements related to the particle current $I$ in the circular tube would not be clearly visible even if we assume that each disk is visible. 
It is also suggested in Fig. \ref{Fig12ParCurEneRatio} that the ratio $K_{cur}/K_{tot}$ takes its maximum value at finite values of $T_{2}/T_{1}$, and is symmetric under the change of the sign of $\ln (T_{2}/T_{1})$.

\begin{figure}[!t]
\vspfigA
\begin{center}
\includegraphics[width=\widthfig]{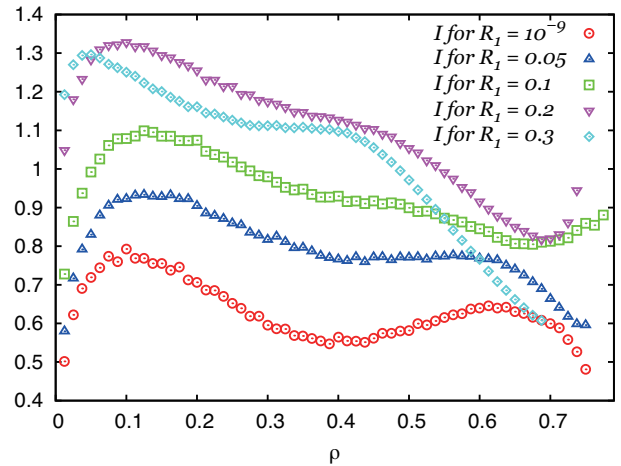}
\vspfigB
\caption{
The particle currents $I$ as functions of the particle density $\rho$ in the cases of $k_{B}T_{2}=2$, $\varphi = \pi/4$, and $\phi = 0$, and $R_{1}=10^{-9}$ (red circles), $R_{1}=0.05$ (blue triangles), $R_{1}=0.1$ (green squares), $R_{1}=0.2$ (purple inverted triangles), and $R_{1}=0.3$ (light-blue diamonds).  
}
\label{Fig13ParCurDenHigh}
\end{center}
\vspfigC
\end{figure}  
%
Our results show that the circulating particle current $I$ by hard disks in a circular tube with a temperature difference decreases monotonically as the particle density decreases in a low-density region. 
Moreover, they also imply that the particle current $I$ is suppressed in a high-density region, probably because particles tend to be stuck inside the tube and are prevented to flow in such a region. 
(See the graph of $I$ in $\rho > 10^{-2}$ in Fig. \ref{Fig3ParCurDen} as that showing this property.) 
However, the particle current $I$ in a high-density region is not described by a monotonically decreasing function of the particle density $\rho$ in general. 
In order to discuss this point, we show Fig. \ref{Fig13ParCurDenHigh} for the graphs of the particle currents $I$ as functions of the particle density $\rho$ in the cases of $k_{B}T_{2}=2$, $\varphi = \pi/4$, and $\phi = 0$, and $R_{1}=10^{-9}$ (red circles), $R_{1}=0.05$ (blue triangles), $R_{1}=0.1$ (green squares), $R_{1}=0.2$ (purple inverted triangles), and $R_{1}=0.3$ (light-blue diamonds).  
This figure suggests that in the density region of $\rho > 0.2$ the graphs of $I$ have flat regions as functions of $\rho$ in the cases of $R_{1}=0.05, 0.1, 0.2$, and $0.3$. 
It is also shown that the particle currents $I$ are even increasing functions of $\rho$ partly in the cases of $R_{1}=10^{-9}$ around $\rho \approx 0.5$, and $R_{1}=0.1, 0.2$ around $\rho \approx 0.75$ with very high densities. 
The origin of these behaviors might be related to a phase transition of hard-disk systems, but it would have to be clarified by additional results and arguments.

\begin{figure}[!t]
\vspfigA
\begin{center}
\includegraphics[width=\widthfig]{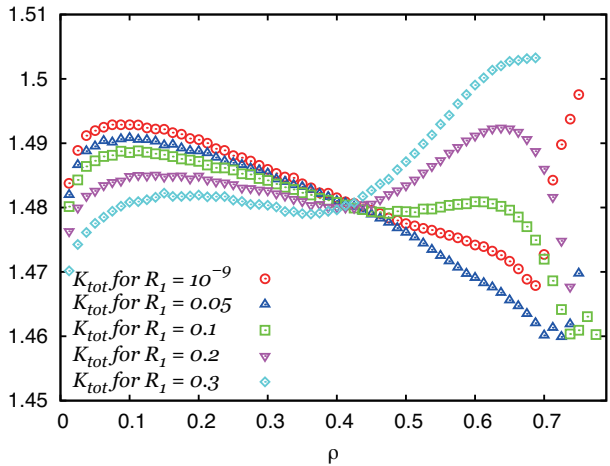}
\vspfigB
\caption{
The average kinetic energies $K_{tot}$ per disk as functions of the particle density $\rho$ in the cases of $k_{B}T_{2}=2$, $\varphi = \pi/4$, and $\phi = 0$, and $R_{1}=10^{-9}$ (red circles), $R_{1}=0.05$ (blue triangles), $R_{1}=0.1$ (green squares), $R_{1}=0.2$ (purple inverted triangles), and $R_{1}=0.3$ (light-blue diamonds). 
}
\label{Fig14KinetEnerg}
\end{center}
\vspfigC
\end{figure}  
%
In this paper, we have discussed transport properties of a particle current and energy currents as nonequilibrium effects caused by a temperature difference. 
On the other hand, our model has many other quantities which are not described as kinds of currents but show important nonequilibrium effects. 
In order to discuss such an example, we show Fig. \ref{Fig14KinetEnerg} for the graphs of the time-average kinetic energies $K_{tot}$ per disk as functions of the particle density $\rho$ in the cases of $k_{B}T_{2}=2$, $\varphi = \pi/4$, and $\phi = 0$, and $R_{1}=10^{-9}$ (red circles), $R_{1}=0.05$ (blue triangles), $R_{1}=0.1$ (green squares), $R_{1}=0.2$ (purple inverted triangles), and $R_{1}=0.3$ (light-blue diamonds). 
This figure shows that the average kinetic energy $K_{tot}$ in our nonequilibrium model depends on the particle density $\rho$. 
Moreover, the average kinetic energy $K_{tot}$ depends on system geometries such as the radius $R_{1}$, as also shown in Fig. \ref{Fig14KinetEnerg}.   
These properties of $K_{tot}$ are very contrastive to the ones in equilibrium states in which velocities of particles are distributed by the Maxwell distribution and the average kinetic energy is independent of the potential energy (so of the particle density $\rho$) and system geometries in a constant temperature.  
It may also be noted that the average kinetic energy $K_{tot}$ satisfies the inequality $k_{B}T_{1} \leq K_{tot} \leq k_{B}T_{2}$ in the data used for this figure.

It is shown that a circulating particle current also occurs in hard disks confined in a single circle wall which is illustrated by removing the inner circle wall with the radius $R_{1}$ in Fig. \ref{Fig1System} \cite{TMS18}. 
(One may regard this system as an angular momentum generator by a temperature difference.) 
As another variation on the circular-tube system, one might introduce a cyclic spatial area by using the periodic boundary conditions. 
In this paper we considered hard-disk systems, but it would be interesting to consider effects of a different type of particle-particle interactions (e.g. the one by the Lennard-Jones potential) in circulating particle currents and energy currents. 
Instead of the thermal boundary conditions used in this paper, one may use another method for coupling to heat reservoirs, such as the methods with thermal random forces in Langevin equations \cite{R89,K92} or thermostat techniques \cite{T99,FS02}.

In general, a temperature difference applied to a many-particle system can cause, not only energy currents, but also a density gradient of particles (as known as the Soret effect) inside the system. 
Besides, the density gradient of particles could be a driving force of particle currents (as known as Fick's law). 
These types of energy and particle currents are subjects of the nonequilibrium thermodynamics \cite{GM84,D14}. 
On the other hand, the nonequilibrium thermodynamics is constructed under some postulates, such as local equilibrium assumptions and small magnitudes of thermodynamics forces.  
Whether these postulates could be justified in our circular-tube model with a temperature difference or not, and how the circulating particle current and energy currents in this model could be discussed by the nonequilibrium thermodynamics or its generalization, still remains as unsettled and important problems.


\end{document}